\documentclass[proceedings]{JHEP} 
\usepackage{epsfig}

\def\particleone{\sl}
\def\tiny{\vrule width 0pt}
\def\star{{\bf *}}

\def\PM{\relax\ifmmode{\pm}\else{$\pm$}\fi}

\def\decays{\relax\ifmmode{\rightarrow}\else{$\rightarrow$}\fi\tiny}

\def\EPEM{\relax\ifmmode{e^+e^-}\else{$e^+e^-$}\fi}
\def\epem{\relax\ifmmode{e^+e^-}\else{$e^+e^-$}\fi}

\def\fmode{\relax\ifmmode{f}\else{$f$}\fi}
\def\fbar{\relax\ifmmode{\bar{f}}\else{$\bar{f}$}\fi}

 \def\N{\relax\ifmmode{\nu}\else{$\nu$}\fi}
 \def\NB{\relax\ifmmode{\overline{\nu}}
	\else{$\overline{\nu}$}\fi}

 \def\NE{\relax\ifmmode{\nu_e}\else{$\nu_e$}\fi}
 \def\NEB{\relax\ifmmode{\overline{\nu}\tiny_e}
	\else{$\overline{\nu}\tiny_e$}\fi}

 \def\E{\relax\ifmmode{e}\else{$e$}\fi}
 \def\EP{\relax\ifmmode{e^+}\else{$e^+$}\fi}
 \def\EM{\relax\ifmmode{e^-}\else{$e^-$}\fi}
 \def\EPM{\relax\ifmmode{e^\pm}\else{$e^\pm$}\fi}
 \def\EMP{\relax\ifmmode{e^\mp}\else{$e^\mp$}\fi}

 \def\NM{\relax\ifmmode{\nu_\mu}\else{$\nu_\mu$}\fi}
 \def\NMB{\relax\ifmmode{\overline{\nu}\tiny_\mu}
	\else{$\overline{\nu}\tiny_\mu$}\fi}

 \def\M{\relax\ifmmode{\mu}\else{$\mu$}\fi}
 \def\MP{\relax\ifmmode{\mu^+}\else{$\mu^+$}\fi}
 \def\MM{\relax\ifmmode{\mu^-}\else{$\mu^-$}\fi}
 \def\MPM{\relax\ifmmode{\mu^\pm}\else{$\mu^\pm$}\fi}
 \def\MMP{\relax\ifmmode{\mu^\mp}\else{$\mu^\mp$}\fi}

 \def\NT{\relax\ifmmode{\nu_\tau}\else{$\nu_\tau$}\fi}
 \def\NTB{\relax\ifmmode{\overline{\nu}\tiny_\tau}
	\else{$\overline{\nu}\tiny_\tau$}\fi}

 \def\T{\relax\ifmmode{\tau}\else{$\tau$}\fi}
 \def\TP{\relax\ifmmode{\tau^+}\else{$\tau^+$}\fi}
 \def\TM{\relax\ifmmode{\tau^-}\else{$\tau^-$}\fi}
 \def\TPM{\relax\ifmmode{\tau^\pm}\else{$\tau^\pm$}\fi}
 \def\TMP{\relax\ifmmode{\tau^\mp}\else{$\tau^\mp$}\fi}

 \def\NL{\relax\ifmmode{\nu_\ell}\else{$\nu_\ell$}\fi}
 \def\NLB{\relax\ifmmode{\overline{\nu}\tiny_\ell}
	\else{$\overline{\nu}\tiny_\ell$}\fi}

 \def\L{\relax\ifmmode{\ell}\else{$\ell$}\fi}
 \def\LP{\relax\ifmmode{\ell^+}\else{$\ell^+$}\fi}
 \def\LM{\relax\ifmmode{\ell^-}\else{$\ell^-$}\fi}
 \def\LPM{\relax\ifmmode{\ell^\pm}\else{$\ell^\pm$}\fi}
 \def\LMP{\relax\ifmmode{\ell^\mp}\else{$\ell^\mp$}\fi}

 \def\PI{\relax\ifmmode{\pi}\else{$\pi$}\fi}
 \def\PIP{\relax\ifmmode{\pi^+}\else{$\pi^+$}\fi}
 \def\PIZ{\relax\ifmmode{\pi^0}\else{$\pi^0$}\fi}
 \def\PIM{\relax\ifmmode{\pi^-}\else{$\pi^-$}\fi}
 \def\PIPM{\relax\ifmmode{\pi^\pm}\else{$\pi^\pm$}\fi}
 \def\PIMP{\relax\ifmmode{\pi^\mp}\else{$\pi^\mp$}\fi}
 \def\PIPMZ{\relax\ifmmode{\pi^{\pm,0}}\else{$\pi^{\pm,0}$}\fi}

 \def\K{\relax\ifmmode{K}\else{{\particleone K}}\fi}
 \def\KB{\relax\ifmmode{\overline{K}}
	\else{$\overline{{\particleone K}}$}\fi}

 \def\KZ{\relax\ifmmode{K^0}\else{{\particleone K}$^0$}\fi}
 \def\KSH{\relax\ifmmode{K^0_S}\else{{\particleone K}$^0_S$}\fi}
 \def\KLO{\relax\ifmmode{K}^0_L\else{{\particleone K}$^0_L$}\fi}
 \def\KZB{\relax\ifmmode{\overline{K}\tiny^0}
	\else{$\overline{{\particleone K}}\tiny^0$}\fi}

 \def\KP{\relax\ifmmode{K^+}\else{{\particleone K}$^+$}\fi}
 \def\KM{\relax\ifmmode{K^-}\else{{\particleone K}$^-$}\fi}
 \def\KPM{\relax\ifmmode{K^\pm}\else{{\particleone K}$^\pm$}\fi}
 \def\KMP{\relax\ifmmode{K^\mp}\else{{\particleone K}$^\mp$}\fi}

\def\D{\relax\ifmmode{{D}}\else{{\particleone D}}\fi\tiny}
\def\DB{\relax\ifmmode{\overline{{D}}}
	\else{$\overline{{D}}$}\fi}

\def\DZ{\relax\ifmmode{{D}^0}\else{{\particleone D}$^0$}\fi}
\def\DZB{\relax\ifmmode{\overline{{D}}\tiny^0}
	\else{$\overline{{D}}\tiny^0$}\fi}

\def\DP{\relax\ifmmode{{D}^+}\else{{\particleone D}$^+$}\fi}
\def\DM{\relax\ifmmode{{D}^-}\else{{\particleone D}$^-$}\fi}
\def\DPM{\relax\ifmmode{{D}^\pm}\else{{\particleone D}$^\pm$}\fi}
\def\DMP{\relax\ifmmode{{D}^\mp}\else{{\particleone D}$^\mp$}\fi}

\def\Ds{\relax\ifmmode{{D}_s}\else{{\particleone D}$_s$}\fi}
\def\DsP{\relax\ifmmode{{D}_s^+}\else{{\particleone D}$_s^+$}\fi}
\def\DsM{\relax\ifmmode{{D}_s^-}\else{{\particleone D}$_s^-$}\fi}
\def\DsPM{\relax\ifmmode{{D}_s^\pm}\else{{\particleone D}$_s^\pm$}\fi}
\def\DsMP{\relax\ifmmode{{D}_s^\mp}\else{{\particleone D}$_s^\mp$}\fi}

\def\KS{\relax\ifmmode{K^\star}\else{{\particleone K}$^\star$}\fi}
\def\KSB{\relax\ifmmode{\overline{K}\tiny^\star}
	\else{$\overline{{\particleone K}}\tiny^\star$}\fi}

\def\KSZ{\relax\ifmmode{K^{\star0}}\else{{\particleone K}$^{\star0}$}\fi}
\def\KSZB{\relax\ifmmode{\overline{K}\tiny^{\star0}}
	\else{$\overline{{\particleone K}}\tiny^{\star0}$}\fi}

\def\KSP{\relax\ifmmode{K^{\star+}}\else{{\particleone K}$^{\star+}$}\fi}
\def\KSM{\relax\ifmmode{K^{\star-}}\else{{\particleone K}$^{\star-}$}\fi}
\def\KSPM{\relax\ifmmode{K^{\star\pm}}\else{{\particleone K}$^{\star\pm}$}\fi}
\def\KSMP{\relax\ifmmode{K^{\star\mp}}\else{{\particleone K}$^{\star\mp}$}\fi}

\def\DS{\relax\ifmmode{D^\star}\else{{\particleone D}$^\star$}\fi}
\def\DSB{\relax\ifmmode{\overline{D}\tiny^\star}
	\else{$\overline{{\particleone D}}\tiny^\star$}\fi}

\def\DSZ{\relax\ifmmode{D^{\star0}}\else{{\particleone D}$^{\star0}$}\fi}
\def\DSZB{\relax\ifmmode{\overline{D}\tiny^{\star0}}
	\else{$\overline{{\particleone D}}\tiny^{\star0}$}\fi}

\def\DSP{\relax\ifmmode{D^{\star+}}\else{{\particleone D}$^{\star+}$}\fi}
\def\DSM{\relax\ifmmode{D^{\star-}}\else{{\particleone D}$^{\star-}$}\fi}
\def\DSPM{\relax\ifmmode{D^{\star\pm}}\else{{\particleone D}$^{\star\pm}$}\fi}
\def\DSMP{\relax\ifmmode{D^{\star\mp}}\else{{\particleone D}$^{\star\mp}$}\fi}

\conference{Heavy Flavours 8, Southampton, UK, 1999}
\title{Charm Mixing and Rare Decays}
\author{Paul D. Sheldon\\Department of Physics and Astronomy, Vanderbilt 
University, Nashville, TN, USA, 37235 \\ 
Email: \email{paul.sheldon@vanderbilt.edu}}
\abstract{
There has been significant recent experimental activity on the 
related topics of 
charm mixing and rare (flavor changing neutral current) decay.
For mixing, 
several new results 
from direct (wrong sign) searches   
and first results from lifetime difference 
($\Delta \Gamma$) searches have been reported.
Limits for $r_{\rm mix}$ of approximately $5\times10^{-4}$ 
(or better!) are possible 
from work in progress.  For rare decays, 
sensitivities to branching ratios are now 
at the level of a few $\times$ $10^{-6}$.
}

\begin{document}

It was 
the absence of flavor changing neutral currents (FCNC) that led 
Glashow, Iliopoulos, and Maiani (GIM) to propose a suppression 
mechanism which 
required the existence of
a fourth quark (charm) in 1970~\cite{GIM}.  
GIM suppression works very well for charm:  the short-distance
standard model predictions for FCNC decays and mixing are extremely
small.
It is this suppression that makes searching for charm mixing
and rare decays interesting:  the distinctive signatures of these effects 
and the very small standard model expectations provide an opportunity 
to search for new physics.


\section{Indirect Searches {\normalsize for} New Physics}


As a reminder of why GIM suppression works so well for
charm, consider the box diagrams which 
represent the lowest order short-distance contribution
to \DZ--\DZB\ 
mixing (figure \ref{mixing_fd}).  
The mixing amplitude calculated from 
these diagrams is proportional to~\cite{Pich}:
%
%
%
\begin{equation} \langle \DZB | H_{\rm wk} | \DZ \rangle \propto 
\!\!\!\!\sum_{i,j=d,s,b}\!\!\!\!V_{ci}^{\star}\ V_{ui}^{ }\ 
V_{cj}^{ }\ V_{uj}^{\star }\ S(m_i^2,m_j^2) 
\label{unimix}\end{equation}
\noindent where $V_{ij}$ are the 
Cabibbo-Kobayashi-Maskawa (CKM) matrix elements.
If the quark masses $m_i$ were all equal,
the loop functions $S(m_i^2,m_j^2)$ would all be equal and  
the amplitude would be zero due to the 
unitarity 
($\sum V_{ci}^{\star}\ V_{ui}^{ } = 0$)   
of the CKM matrix.  
If the mass differences are small, GIM very nearly works and 
mixing is small. 
\FIGURE{\epsfig{file=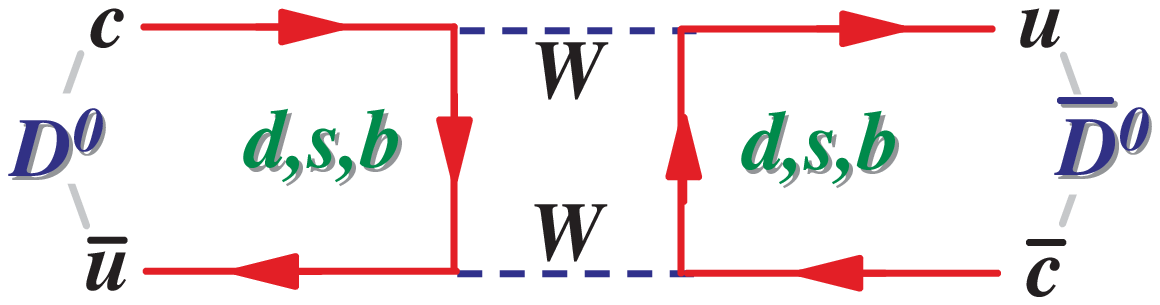,width=5.5cm}
\caption{One of the two box diagram for mixing. Swapping the internal
$W$ and quark lines gives the other.}\label{mixing_fd}}
For charm, the CKM factor 
$V_{cb}^{\star}\ V_{ub}^{ }$ 
is insignificant
($\sim \lambda^5$ in the
Wolfenstein parameterization~\cite{Wolf1}) 
relative to the factors   
$V_{cd}^{\star}\ V_{ud}^{ }$ and $V_{cs}^{\star}\ V_{us}^{ }$
(both $\sim \lambda$).  Only the $i,j = d,s$ terms contribute
in equation \ref{unimix},  
and the mass difference between the $d$ and $s$  
quarks is relatively small.
The \DZ--\DZB\  
mixing probability calculated from these box diagrams is 
$\sim 10^{-10} - 10^{-9}$~\cite{Datta}.
By contrast, neutral $b$ quark mesons exhibit large mixing because
the top mass is so large.  For example, for $B_d^0$ 
the three CKM factors ($V_{ib}^{\star}\ V_{id}^{ }$, $i=u,c,t$)
are roughly equal ($\sim A \lambda^3$) so the contribution
of the top quark dominates.  

Because the short distance predictions for charm mixing 
are so small, long distance contributions may be important.
They are more difficult to calculate, and there is
significant disagreement over their 
size~\cite{Wolfenstein,Georgi,Pakvasa}.  With long distance effects,
$r_{\rm mix}$ may be as large as $10^{-3}$.

FCNC decays are also suppressed by the GIM mechanism.  Even with
long distance effects, the expected branching ratios 
for rare charm decays are in the range 
$10^{-19} - 10^{-6}$~\cite{Pakvasa,Schwartz}.

If there is new physics such as a fourth generation of quarks, leptoquarks,
etc., it can contribute to the box diagrams for mixing or the penguin
diagrams for FCNC decays~\cite{Hewett}.  Because the standard model 
predictions are so small, there is a large window to observe the
additional contributions of this new physics, unhidden by standard
model effects.  Using high statistics instead of high energy, 
one indirectly probes for new particle
states which cannot be produced directly.


\section{Charm Mixing 101}

There are enough differences in the 
``nuts and bolts'' of mixing for charm and beauty that it is useful to 
review the important points for charm.

\subsection{Basic Charm Mixology}

The neutral \D\ mesons evolve according to:
\begin{eqnarray*} 
i {\partial\over{\partial t}} 
\left( \begin{array}{cc} \DZ \\ \DZB \end{array} \right) = 
H_{\rm wk} \left( \begin{array}{cc} \DZ \\ \DZB \end{array} \right)
\hspace*{3.3cm} \\
\hspace*{1.0cm}
= \left( \begin{array}{cc}
M - i \Gamma/2 & M_{12} - i \Gamma_{12}/2 \\
M_{12}^\star - i \Gamma_{12}^\star/2 & M - i \Gamma/2 
\end{array} \right) 
\left( \begin{array}{cc} \DZ \\ \DZB \end{array} \right).
\end{eqnarray*}
\noindent Diagonalizing gives weak eigenstates 
\begin{displaymath}
D_H = p\DZ + q\DZB;\qquad D_L = p\DZ - q\DZB
\end{displaymath}
\begin{displaymath}
{q\over p} = \left[
{{M_{12}^\star - i \Gamma_{12}^\star/2}\over{M_{12} - i \Gamma_{12}/2}}
\right]^{1/2} 
\end{displaymath}
of definite mass and lifetime:
\begin{displaymath}
M_{H,L} = 
M \pm \Re[(M_{12}^\star - i \Gamma_{12}^\star/2)
(M_{12} - i \Gamma_{12}/2)]^{1/2} 
\end{displaymath}
\begin{displaymath}
\Gamma_{H,L} = 
\Gamma \mp 2\Im[(M_{12}^\star - i \Gamma_{12}^\star/2)
(M_{12} - i \Gamma_{12}/2)]^{1/2}.
\end{displaymath}
\noindent These eigenstates evolve with time as:
\begin{displaymath}
D_{H,L}(t) = e^{-iM_{H,L}t-{1\over2}\Gamma_{H,L}t} D_{H,L}(0). 
\end{displaymath}
\noindent If $H_{\rm wk}$ conserves CP, then $D_H$ and $D_L$ are CP
eigenstates and $p = q = 1$.

If you start with a \DZ, the probability that it is a \DZB\ at time
$t$ is:
\begin{eqnarray*}
r_{\rm mix}(t) = \Gamma(\DZ\decays\DZB)\hspace{4.0cm} \\
\hspace{0.8cm} = {\textstyle{1\over 4}} \left| {\textstyle {q\over p}}\right|^2
\left[ e^{-\Gamma_H t} + e^{-\Gamma_L t} - 2 e^{-\Gamma t} 
\cos \Delta M t \right]
\end{eqnarray*}
where $\Delta M = (M_H - M_L)$ and $\Delta \Gamma = (\Gamma_H - \Gamma_L)$.

Experimental limits 
on \DZ--\DZB\ mixing~\cite{e615,e691,e791SL,e791had,alephws}  
indicate that  
$\Delta M << \Gamma$ and $\Delta \Gamma << \Gamma$, so
\begin{displaymath}
r_{\rm mix}(t) = {\textstyle {1\over 4}} e^{-\Gamma t}
\left| {\textstyle {q\over p}}\right|^2 \left( \Delta M^2 + 
{\textstyle {1\over 4}} \Delta \Gamma^2 \right) t^2.
\end{displaymath}
If we define $x = \Delta M / \Gamma$ and $y = \Delta \Gamma / 2 \Gamma$, then
\begin{equation}
r_{\rm mix}(t) = {\textstyle {1\over 4}} e^{-\Gamma t}
\left| {\textstyle {q\over p}}\right|^2 \left( x^2 + y^2 \right) 
(\Gamma^2 t^2).
\label{SLmix}\end{equation}
Integrated over all time:  
\begin{equation}
r_{\rm mix} = {1\over 2} 
\left| {q\over p}\right|^2 (x^2 + y^2).
\label{rmixint}\end{equation}
For the charge conjugate process ($\DZB\decays\DZ$):
\begin{equation}
\overline{r}_{\rm mix}(t) = {\textstyle {1\over 4}} e^{-\Gamma t}
\left| {\textstyle {p\over q}}\right|^2 
\left( x^2 + y^2 \right) 
(\Gamma^2 t^2)
\label{SLmixBar}\end{equation}
so that $r_{\rm mix} = \overline{r}_{\rm mix}$ only if 
$|q / p| = 1$.

Expectations from  
standard model short distance calculations are that $x$ and $y$ are
approximately equal.  However, if
$r_{\rm mix}$ is large, the source is likely to be $x$.  While long
distance effects and/or new physics can increase $\Delta M$ substantially, 
they do not make significant contributions to 
$\Delta \Gamma$~\cite{Wolfenstein,Hewett,qoverp}. 

\subsection{Search Strategies}

There are two basic methods currently employed to search for charm mixing.
In direct or ``wrong sign'' searches, one looks for 
$\DZ \decays \DZB \decays \fbar$, where 
$\fbar$ is a Cabibbo favored (CF) mode such as $\KP\PIM$ or $\KP\MM\NM$.  
The sign of the daughter \K\
or \M\ distinguishes \DZB\decays\fbar\ from 
\DZ\decays\fmode.  The produced \D\ is
identified using a \DS\ tag: 
$\DSP \decays \DZ\PIP$.  The sign of the ``bachelor'' pion from the
\DS\ decay tags the produced neutral \D\ as a \DZ\ (\PIP) or 
\DZB\ (\PIM).    The signal for mixing is that 
the bachelor pion and the \K\ daughter have the same charge:
$\DSP\decays\PIP\DZ;\ 
\DZ\decays\DZB\decays\KP\PIM$.

The second method is to look for a lifetime difference between $D_H$ and
$D_L$.  If $D_H$ and $D_L$ are CP$^+$ and
CP$^-$ eigenstates, then:
\begin{equation}
y_{\rm CP} \equiv  {{\T_- - \T_+}\over{\T_- + \T_+}} =
{\Delta\Gamma\over{2 \Gamma}} = y.
\label{ydiff}\end{equation} 
The lifetimes $\T_+$ and $\T_-$ can be measured using neutral \D\ 
decays to states of definite CP, such as 
$\DZ\decays\KP\KM$ (CP$^+$) or $\DZ\decays\KSH\phi$ (CP$^-$).  
Even if one relaxes the ``no CP violation'' requirement, 
$y_{\rm CP} \approx y$ 
because 
it is known~\cite{E687cp} that CP violation is small in
charm decay and therefore $D_H$ and $D_L$ are approximately CP eigenstates. 

\subsection{The DCS Wrinkle}\label{wrinkle}

For wrong sign hadronic searches 
($\fbar = \KP\PIM$, $\KP\PIM\PIP\PIM$,...)
there is an interesting ``wrinkle''.
Because mixing is at best a small effect, 
doubly Cabibbo suppressed (DCS) decay cannot be ignored when looking for 
a wrong-sign signal~\cite{blaylock,Browder}.  Via DCS, 
$\DZ\decays\fbar$ can occur directly and not just through mixing
$\DZ\decays\DZB\decays\fbar$.  
In this case:
\begin{displaymath}
A_{\rm WS} = A_{\rm DCS}(\DZ\decays\fbar) + 
A_{\rm mix}(\DZ\decays\DZB\decays\fbar)
\end{displaymath}
and the wrong sign decay rate becomes:
\begin{eqnarray}
r_{\rm WS} = 
e^{-\Gamma t} \left| {\textstyle {q\over p}}\right|^2
\left\langle\fbar\left|H\right|\DZB\right\rangle^2_{\rm CF} 
\times \left\{ |\lambda|^2 + \nonumber \right. \hspace*{1.0cm} \\
\left.\textstyle{1\over4} (x^2 + y^2) \Gamma^2 t^2 
+ \left(\Re(\lambda)y + \Im(\lambda)x\right) \Gamma t \right\}\hspace*{0.7cm} 
\label{hadmix}\end{eqnarray}
where
\begin{displaymath}
\lambda = 
\textstyle{{p\left\langle\fbar|H|\DZ\right\rangle_{\rm DCS}}\over
{q\left\langle\fbar|H|\DZB\right\rangle_{\rm CF}}};\quad
\overline{\lambda} = 
\textstyle{{q\left\langle\fmode|H|\DZB\right\rangle_{\rm DCS}}\over
{p\left\langle\fmode|H|\DZ\right\rangle_{\rm CF}}}.
\end{displaymath}
By measuring the proper time of the decay of the neutral $D$ meson, one can 
disentangle the contributions from DCS
and mixing:  while all terms in equation \ref{hadmix} 
share a common exponential time dependence,
the mixing term ($x^2 + y^2$) is proportional to $t^2$, 
the DCS term ($|\lambda|^2$) has no additional time dependence, and
the interference term is proportional to $t$.

The interference term is interesting:  it may be observable
even if the mixing term isn't (in which case 
mixing would be observable in 
hadronic modes but not semileptonic!).  Alternatively, if it has
the opposite sign of the mixing term (destructive interference) 
and is of roughly equal size 
the effects of mixing could be masked~\cite{blaylock}.  
Finally, this term could allow
experimenters to distinguish between $\Delta \Gamma$ and $\Delta M$
contributions to mixing.  This is especially true if CP violation 
in mixing is very small or zero (see section \ref{simplify}).

Note also that the CP conjugate rate for wrong sign decays is not
the same as equation \ref{hadmix}:
\begin{eqnarray}
\overline{r}_{\rm WS} = 
e^{-\Gamma t} \left| {\textstyle {p\over q}}\right|^2
\left\langle\fmode\left|H\right|\DZ\right\rangle^2_{\rm CF} 
\times \left\{ |\overline{\lambda}|^2 + \nonumber \right.\hspace*{1.0cm} \\
\left.\textstyle{1\over4} (x^2 + y^2) \Gamma^2 t^2 
+ \left(\Re(\overline{\lambda})y + \Im(\overline{\lambda})x\right) 
\Gamma t \right\}. \hspace*{0.7cm}
\label{hadmixbar}\end{eqnarray}

\subsection{Simplifying Assumptions}\label{simplify}

The wrong sign rates of equations \ref{hadmix} and \ref{hadmixbar} 
can be simplified
by making assumptions about the nature of CP violation.  For example, it 
is likely that 
there is no direct CP violation in 
CF or 
DCS decays\footnote{For direct CP violation to appear in a decay mode,
there must be two amplitudes that contribute significantly 
to that final state.} 
and that CP 
is not violated in charm mixing~\cite{qoverp},  
{\it i.e.} 
\begin{equation}
|q / p| = 1.
\label{ass1}\end{equation} 
 If we assume both of the above 
then 
\begin{equation}
|\lambda| = |\overline{\lambda}|.
\label{ass2}\end{equation}
Defining the strong (final state interaction) phase $\delta$ and the
CP violating phase $\phi$:
\begin{eqnarray}
\sqrt{R_{\rm DCS}} \cdot e^{i\delta} & = & 
\textstyle{
{\left\langle\fbar\left|H\right|\DZ\right\rangle_{\rm DCS}} \over 
{\left\langle\fbar\left|H\right|\DZB\right\rangle_{\rm CF}}} \nonumber \\
e^{i\phi} & = & \textstyle{p \over q}
\label{angledefs}\end{eqnarray}
the wrong sign rates become:
\begin{eqnarray}
r_{\rm WS},\ \overline{r}_{\rm WS} =
e^{-\Gamma t} 
\left\langle\fbar\left|H\right|\DZB\right\rangle^2_{\rm CF} \hspace*{2.3cm}
\nonumber \\
\times \left\{ R_{\rm DCS} + 
\textstyle{1\over2} r_{\rm mix} \Gamma^2 t^2 + \hspace*{1.5cm}
\nonumber \right.  \\
\mbox{} + [ y \sqrt{R_{\rm DCS}}\: \cos(\delta\pm\phi) \hspace*{1.3cm} 
\nonumber \\
\left. - x \sqrt{R_{\rm DCS}}\: \sin(\delta\pm\phi)]\: \Gamma t \right\}. 
\hspace*{0.7cm}
\label{hadmixass1}\end{eqnarray}

If we assume CP invariance ($\phi = 0$) then 
$r_{\rm WS} = \overline{r}_{\rm WS}$ and:
\begin{eqnarray}
r_{\rm WS} = 
e^{-\Gamma t} 
\left\langle\fbar\left|H\right|\DZB\right\rangle^2_{\rm CF} \hspace*{3.2cm}
\nonumber \\
\times \left\{ R_{\rm DCS} + 
\textstyle{1\over2} r_{\rm mix} \Gamma^2 t^2 + 
y' \sqrt{R_{\rm DCS}}\: \Gamma t \right\} \hspace*{0.8cm}
\label{hadmixass2}\end{eqnarray}
where
\begin{equation}
y' = y\cos\delta - x\sin\delta;\ \ 
x' = x\cos\delta + y\sin\delta.
\end{equation}
If $\delta$ is small, as has been argued~\cite{Browder,smalldelta}, 
then  
$y' \decays y$ and $x' \decays x$.

If one has 
sufficient proper time resolution, 
equation \ref{hadmixass2} shows that the contributions  
due to $x'$ and $y'$ can
be discriminated (assuming CP invariance).

\subsection{Comments on Assumptions}\label{comments}

There may be excellent theoretical motivation for the assumptions made in 
subsection \ref{simplify}.  But, as demonstrated by E791 
in their hadronic wrong sign search~\cite{e791had}, there are no 
technical reasons for experimenters to make them.
E791 first quoted mixing limits based on equations 
\ref{hadmix} and \ref{hadmixbar}
(minimal assumptions).  They then quoted limits after making the simplifying
assumptions of equations 
\ref{ass1} and \ref{ass2}, and finally for the case
of no mixing (DCS only).  This approach avoids two difficulties.  First, 
if experimenters 
quote results based on only one set of assumptions, it can be 
difficult and misleading to compare results from different experiments
(if past history is any guide, they are unlikely to make similar ones).
Second, and more
importantly, assumptions mask possibly interesting phenomena 
(assuming CP invariance 
precludes searching for CP 
violation in mixing).  

The situation in 
$\Delta \Gamma$ searches is slightly different.  There, one 
measures $y_{\rm CP}$,
and $y = y_{\rm CP}$ only if $D_H$ and $D_L$ are CP eigenstates.  
An observation of 
$y_{\rm CP} \neq 0$ is evidence for mixing but extracting $y$ from this 
requires additional information. 

\section{Wrong Sign Mixing Searches}

E791 and ALEPH have recently published results from wrong sign
searches.  
CLEO and FOCUS have shown preliminary results from analyses in 
progress.  
E791 is the only one to present results in the most general case.

\subsection{E791 Wrong Sign Searches}

E791 at Fermilab is a fixed-target hadroproduction experiment.  Using
a 500 GeV \PIM\ beam and thin target foils, they logged $2\times10^{10}$ 
hadronic interactions.  Their spectrometer employed silicon microstrip
detectors for vertexing, two threshold Cerenkov detectors for
particle identification, a muon hodoscope and a lead/liquid scintillator 
electromagnetic detector
for electron identification.  They have published results for wrong
sign searches using hadronic and semileptonic
decay modes.

Their hadronic analysis~\cite{e791had} uses the 
CF decay modes 
$\DZ\decays\KM\PIP\PIM\PIP$
and 
$\DZ\decays\KM\PIP$.
\FIGURE{\epsfig{file=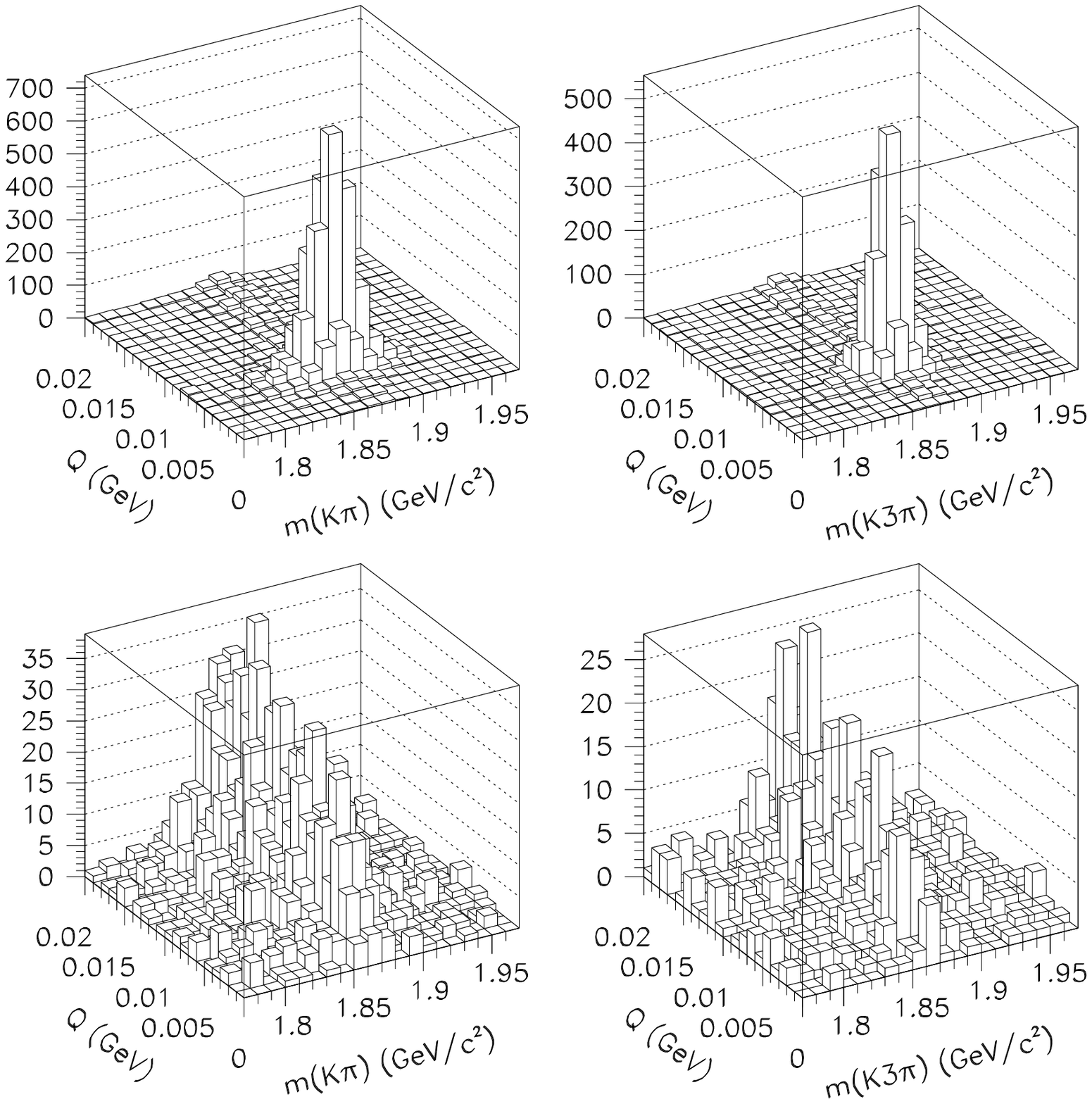,width=7.0cm}
\caption{E791 right sign (top row) and 
wrong sign (bottom row) hadronic data samples.  Candidates for the decay modes $K\pi$ 
and $K3\pi$ are shown in the left and right columns, respectively. 
\DZ\ and \DZB\ candidates are kept separate in the analysis, but are 
combined here.  
$Q = m(\pi K {\rm n} \pi) - m(K {\rm n} \pi) - m(\pi)$.}\label{e791_hmix1}}
Figure
\ref{e791_hmix1} shows the right sign and wrong sign signals they obtain,
where they have combined \DZ\ and \DZB\ modes for the purpose of making the
figure.  In the plots, $Q = m(\pi K{\rm n} \pi) - m(K {\rm n} \pi) - m(\pi)$.
There are 5643 and 3469 reconstructed signal 
events in the right sign $K\pi$ and $K3\pi$ samples.  
In their analysis, they keep the \DZ\ and \DZB\ samples separate, and perform
a simultaneous 
binned maximum likelihood fit to each of the eight resulting data sets.
The \DZ\ mass and width and the $Q$ peak and width are 
constrained to be the same in each data set, but most parameters 
(such as backgrounds) are uncoupled, leading
to a 41 parameter fit in the most general case.

Making no assumptions about CP violation or mixing, 
E791 sets the following 90\% CL limits:
\begin{eqnarray}
r_{\rm mix}(\DZ\decays\DZB) &<& 1.45\%
\nonumber \\
\overline{r}_{\rm mix}(\DZB\decays\DZ) &<& 0.74\%.
\label{e791hadmix0}\end{eqnarray}
\noindent Assuming CP violation only in the interference term 
(see equations \ref{ass1} and \ref{ass2}), they find a 90\% CL limit of:
\begin{equation}
r_{\rm mix} < 0.85\%.
\label{e791hadmix1}\end{equation}
\FIGURE{\epsfig{file=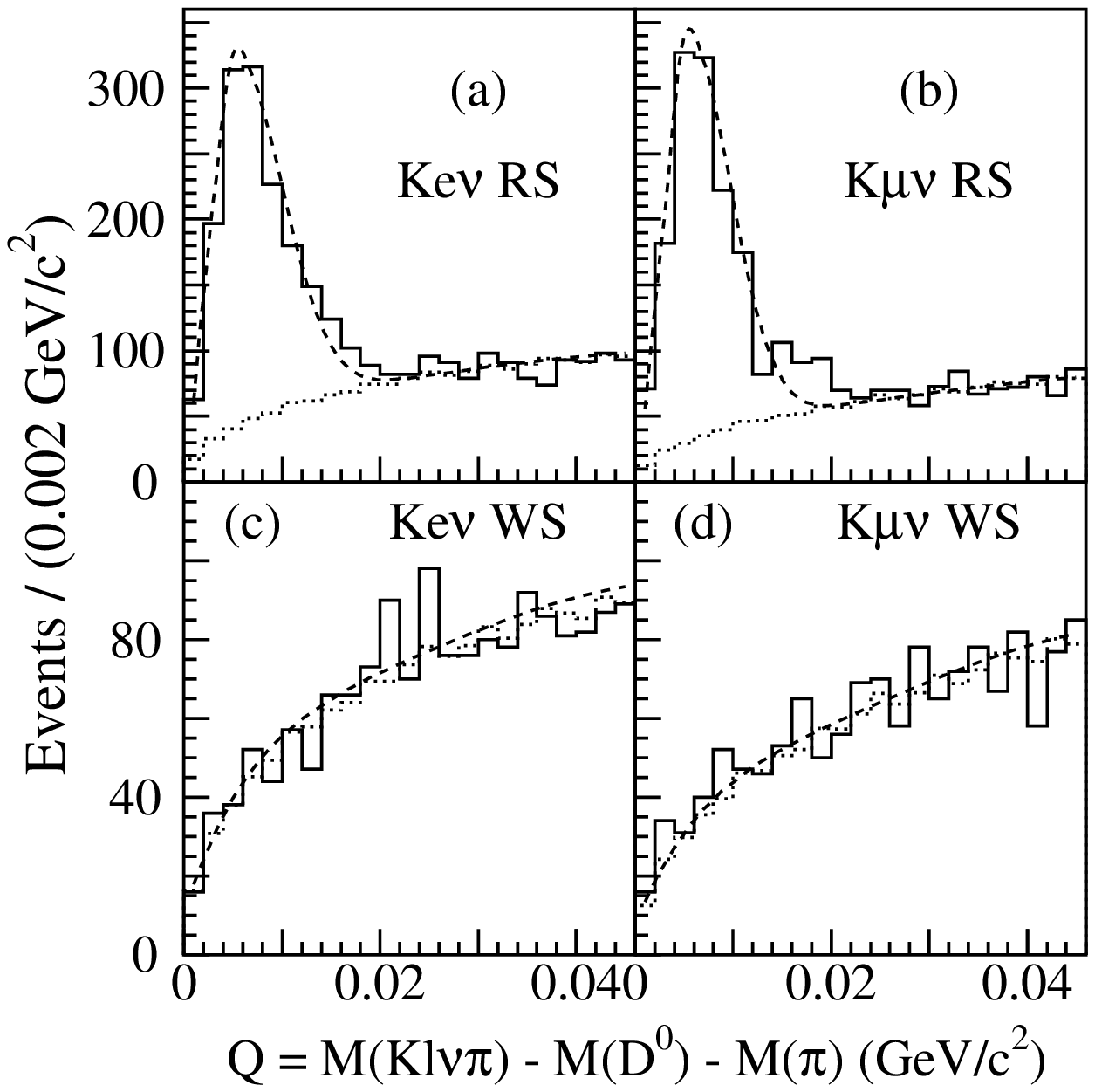,width=5.8cm}
\caption{E791 right sign (top row) and 
wrong sign (bottom row) semileptonic data samples. 
Candidates for the decay modes 
$K e \nu$ and $K \mu \nu$ are shown in the left and 
right columns, respectively.   
}\label{e791_lmix1}}
\noindent If they assume no mixing, they find relative branching
ratios (equation \ref{angledefs}) for DCS modes:
\begin{eqnarray}
R_{\rm DCS}(K\pi) & = & (0.68^{+0.34}_{-0.33}\pm0.07)\% \nonumber \\
R_{\rm DCS}(K3\pi) & = & (0.25^{+0.36}_{-0.34}\pm0.03)\%. 
\label{e791DCSD}\end{eqnarray}
%

The E791 semileptonic wrong sign search~\cite{e791SL} 
uses the \DZ\ decay modes 
$\KM\MP\NM$ and $\KM\EP\NE$.  The missing neutrino gives a two-fold
ambiguity in the \DZ\ momentum; based on Monte Carlo studies they pick the 
higher momentum solution.  Fixing the \DZ\ mass, they then fit the $Q$ and
proper time $t$ distributions.  The right and wrong sign $Q$ 
plots are shown in figure \ref{e791_lmix1}.  The dotted lines show the
estimate of the background they get using an event mixing
technique (they combine \DZ\ candidates from one event with a bachelor
pion from another).  
In this analysis, E791 does not quote a ``minimal
assumption'' result, instead they make the assumption of 
equation \ref{ass1} from 
the outset and fit to a time dependence given by equation \ref{SLmix}.  
The number of right sign decays they
get from their fit is $1237\pm45$ ($Ke\nu$) and $1267\pm44$ ($K\mu\nu$).
The mixing limit they obtain is:
\begin{equation}
r_{\rm mix} < 0.50\%.
\label{e791SLmix}\end{equation}

\subsection{ALEPH Wrong Sign Search}

\FIGURE{\epsfig{file=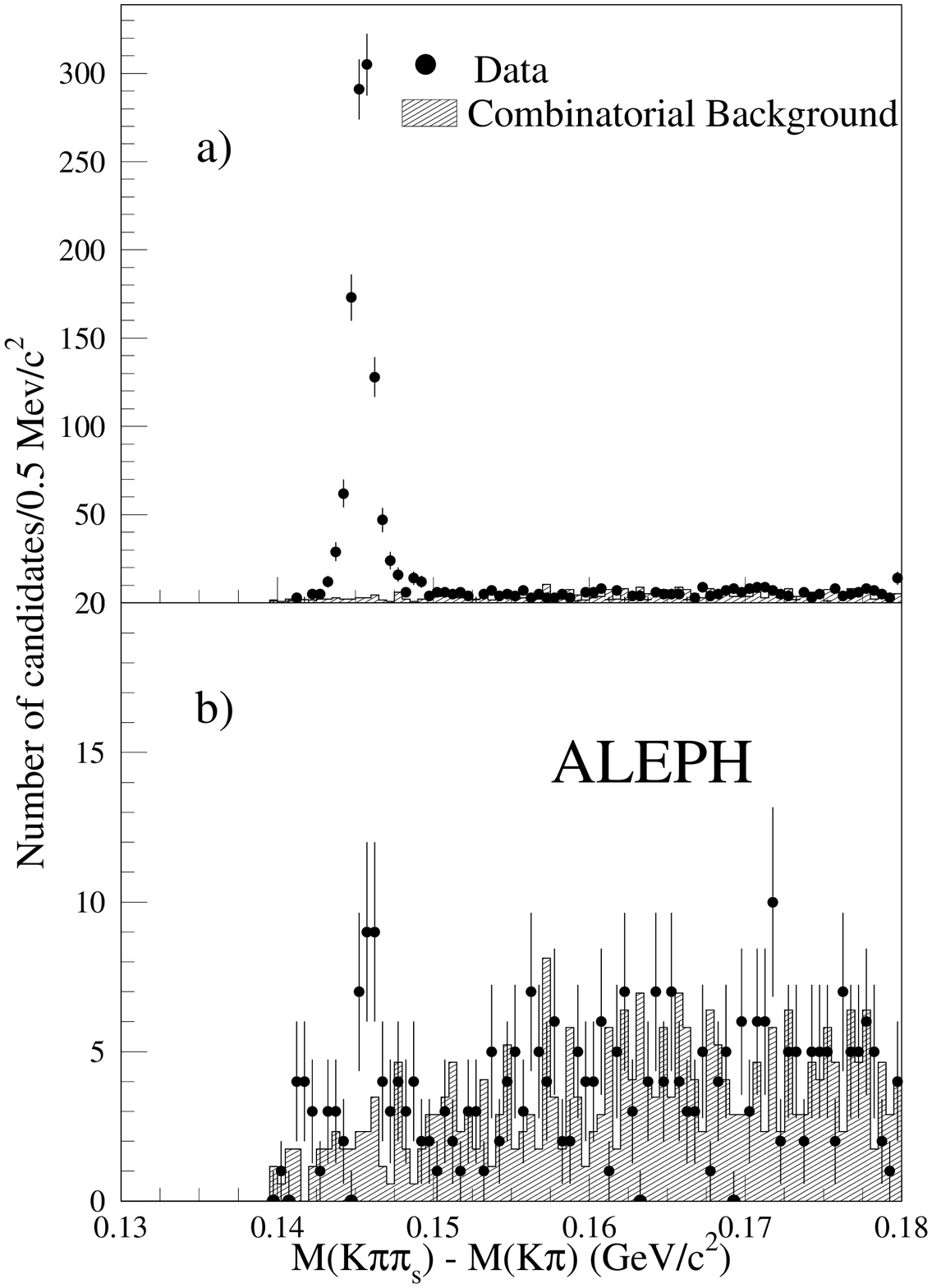,width=5.5cm}
\caption{ALEPH (a) right sign and 
(b) wrong sign data for the $K \pi$ mode.  The dots
with error bars are data and the hatched histogram is the combinatorial
background estimated using sidebands 
in the \DZ\ mass plot.}\label{aleph_mix1}}
The ALEPH collaboration has used its sample of $4\times10^6$ hadronic
$Z$ decays to search for mixing in 
\DZ\decays\KM\PIP~\cite{alephws}.  The right sign
and wrong sign mass plots they obtain are shown in figure 
\ref{aleph_mix1}.  After carefully accounting for combinatoric and
physics\footnote{Backgrounds that come from \DZ\ decays with 
misidentified or missing daughters.} 
backgrounds in these mass plots, they find
$N_{\rm RS} = 1038.8\pm32.5\pm4.3$ and 
$N_{\rm WS} = 19.1\pm6.1\pm3.5$.

To set limits on mixing, they assume CP invariance.  
They parameterize the interference between mixing
and DCS somewhat differently than equation \ref{hadmixass2}.  For the 
proper time dependence of $r_{\rm WS}$ they use:
\begin{equation}
\left\{ R_{\rm DCS} + 
\textstyle{1\over2} r_{\rm mix} \Gamma^2 t^2 + 
\sqrt{R_{\rm DCS}\: r_{\rm mix}}\:\cos \phi\: \Gamma t
\right\}.
\label{alephparam}\end{equation}
They quote 
95\% CL limits on mixing for
three values of $\cos\phi$:
\begin{eqnarray}
r_{\rm mix} < & 0.96\% \quad & \cos\phi = +1 \nonumber \\
r_{\rm mix} < & 0.92\% \quad & \cos\phi = 0 \\
r_{\rm mix} < & 3.6\%\ \quad & \cos\phi = -1. \nonumber
\end{eqnarray}
As noted in section \ref{wrinkle}, the sensitivity to mixing 
decreases dramatically if there is maximal
destructive interference ($\cos\phi = -1$).
It is difficult to decide how to compare these limits to those from
other experiments, given that they fix $\cos\phi$ when extracting 
results and that 
the value of $\cos\phi$ is not known experimentally.

ALEPH used the ratio of observed wrong sign and right sign events to
measure:
\begin{equation}
{{r_{\rm WS}(K\pi)}\over{r_{\rm RS}(K\pi)}} = (1.84\pm0.59\pm0.34)\%.
\label{alephWSresult}\end{equation}
If one makes the assumption of zero mixing, this is also a measurement of
$R_{\rm DCS}(K\pi)$.
ALEPH also quotes 
\begin{displaymath}
R_{\rm DCS}(K\pi) = (1.77^{+0.60}_{-0.56}\pm0.31)\%
\end{displaymath}
which they get from their 
fit to equation \ref{alephparam} in the limit of zero interference
($\cos\phi = 0$) and with the constraint $r_{\rm mix} > 0$.


\subsection{CLEO Wrong Sign Search}

The CLEO collaboration has reported~\cite{CLEOmix} preliminary results
from a wrong sign search for mixing using data from $9.0$ fb$^{-1}$ of 
integrated luminosity taken with the CLEO II.V detector. 
This analysis takes great advantage of the three layer, double-sided 
silicon vertex detector (SVX)
they installed in 1995.  In addition to giving CLEO the ability to 
measure the proper time of charm decays, the SVX greatly improves their
measurement resolution of $Q$, the energy release in the \DSPM\ decay
used to tag the initial state of the neutral $D$.  This improved
resolution enhances their sensitivity to mixing by 
increasing their signal to noise.

After cuts designed to suppress backgrounds from other \DZ\ decays and
``cross-talk'' between their right-sign and wrong-sign samples, 
they obtain the $D$ mass and $Q$ plots for wrong sign candidates shown
in figure~\ref{cleo_ws}.  Superimposed on the data (solid lines) 
in these plots
are colored regions which show the contributions from backgrounds
determined by a two-dimensional fit to $Q$ and $M$.  The background
shapes were determined from Monte Carlo.  From their fit they find 
$54.8\pm10.8$ $\DZ\decays\KP\PIM$ events in their wrong sign sample.
\FIGURE{\epsfig{file=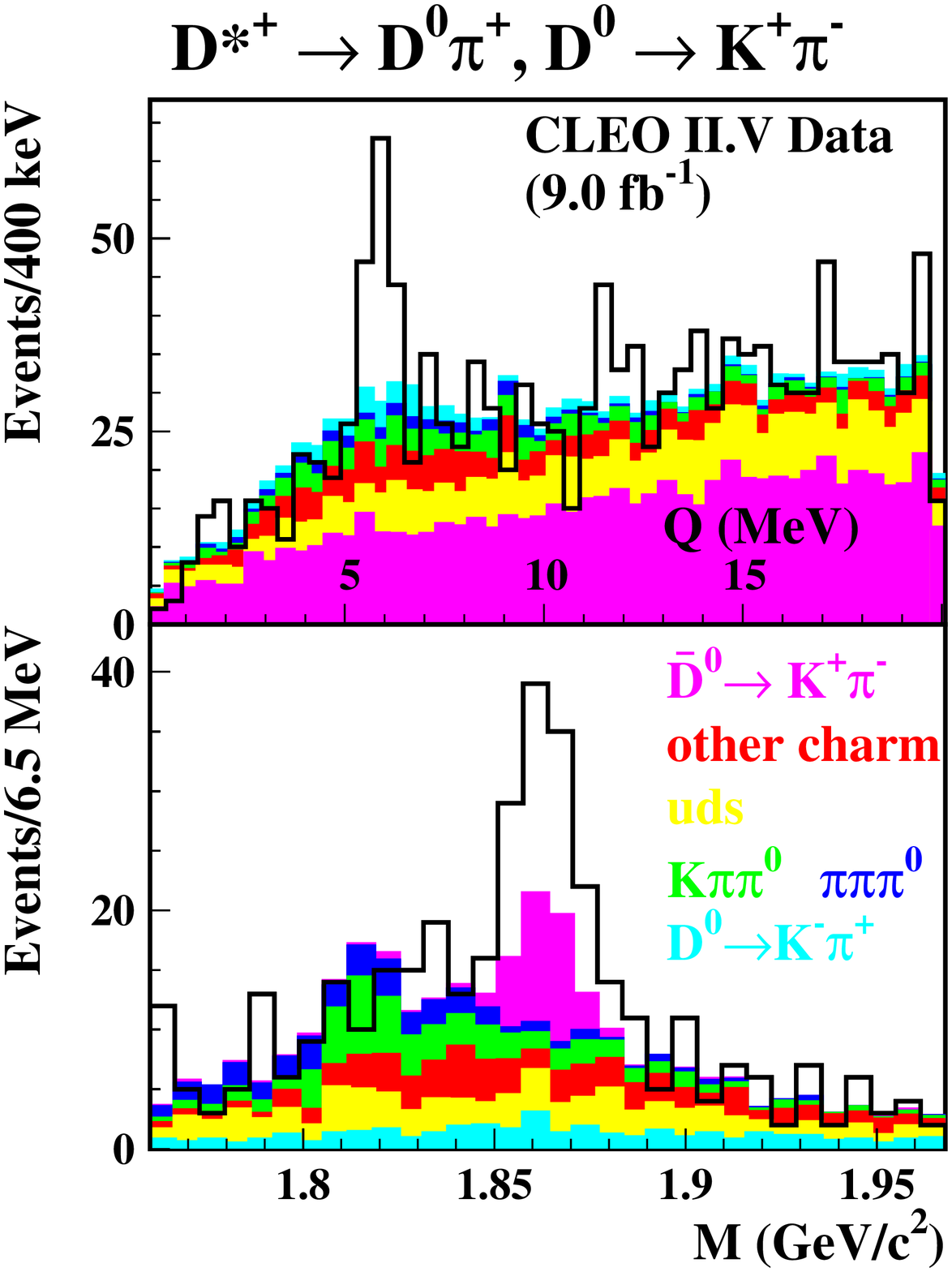,width=5.3cm}
\caption{CLEO wrong sign $Q$ and $M$ distributions
for $K \pi$ candidates from $\DS$ decay. 
Backgrounds from various sources are shown, and come from two-dimensional
fits to $Q$ and $M$ (shapes determined from simulation).}\label{cleo_ws}}
A similar fit to their right sign sample yields $16126\pm126$ 
$\DZ\decays\KM\PIP$ candidates.  Using these numbers, they find
\begin{equation}
{{r_{\rm WS}(K\pi)}\over{r_{\rm RS}(K\pi)}} = (0.34\pm0.07\pm0.06)\%.
\label{cleoWSresult}\end{equation}
Using a time dependent fit to differentiate between mixing and DCS
contributions (described below) they measure
\begin{displaymath}
R_{\rm DCS}(K\pi) = (0.50^{+0.11}_{-0.12}\pm0.08)\%.
\end{displaymath}

CLEO has split its wrong sign sample into \DZ\ and \DZB\ candidates
and states that they see no evidence for a time-integrated CP asymmetry
(the $1\sigma$ statistical error on this asymmetry is 0.19). 
They assume CP invariance and use the proper time distribution given
by (\ref{hadmixass2}) in their fits.  In this parameterization 
the interference term gives independent 
information on $y'$.  From their fits, CLEO finds 
one-dimensional intervals at 95\% CL of:
\begin{equation}
|x'| < 3.2\%;\qquad-5.9\% < y' < 0.3\%.
\end{equation}
Systematic errors are included when finding the above intervals.
It is not possible to determine the sign of 
$x'$ from the fit, which depends only on $(x')^2$.
With this assumption of CP invariance, and if $\delta =0$ 
so that $y' \decays y$ and $x' \decays x$, the above results
give separate 
limits on the $\Delta \Gamma$ and $\Delta M$ contributions
to mixing, an interesting and important result (particularly so given
their excellent sensitivity).  

\FIGURE{\epsfig{file=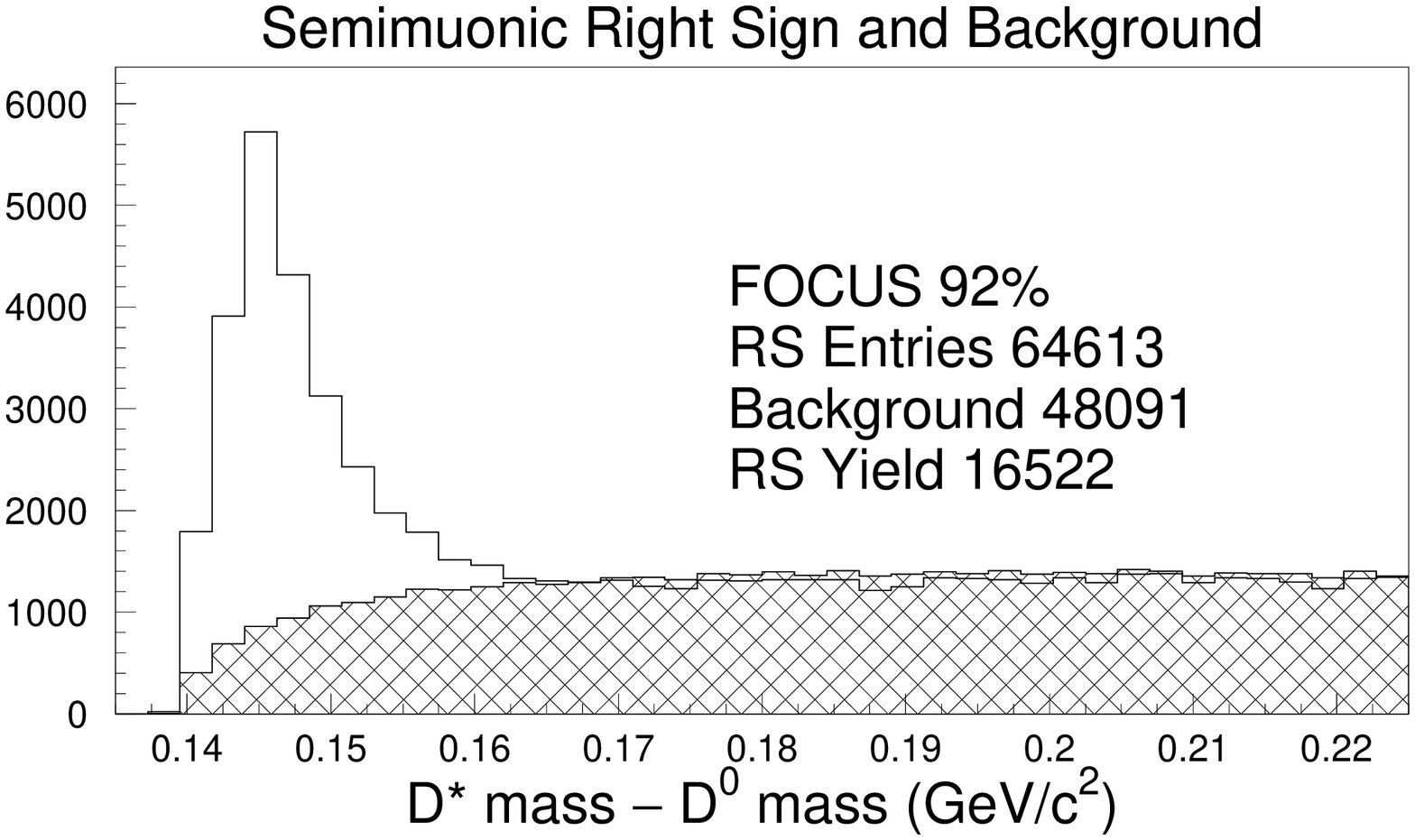,width=6.0cm}
\caption{Preliminary 
FOCUS right sign mass difference
distributions for $K \mu \nu$ candidates from $\DS$ decays.
Combinatoric background (hatched histogram) is estimated using bachelor
pion candidates from other events.
}\label{focus_rs}}

\subsection{FOCUS Wrong Sign Searches}

The Fermilab photoproduction experiment FOCUS collected data in 1996--97.
From this data, it has reconstructed 15 times 
more charm decays than its predecessor E687.  Several improvements were
also made to the spectrometer.  The target was segmented so that more
charm decays occur outside of the target (improves signal to noise).
Vertex position resolution was significantly improved by the addition
of silicon detector planes in the target region. Particle ID was
significantly improved by the addition of a new lead glass
electromagnetic calorimeter, two new muon systems, 
and significant improvements to the Cerenkov reconstruction code.

FOCUS is in the process of performing 
semileptonic and hadronic wrong sign searches.
A mass difference ($\Delta m = m(\DS) - m(\DZ)$) 
plot for their preliminary 
right sign semimuonic 
(\DZ\decays\KM\MP\NM) analysis is shown in 
figure \ref{focus_rs}.  
They model their background shape 
by combining reconstructed \DZ's from one event with soft pions 
from another, and this background is normalized at large $\Delta m$ 
($\Delta m > 0.16$~GeV/c$^2$).  
Using 92\% of their data, they find roughly 16500 background subtracted 
right sign candidates.  
\FIGURE{
\epsfig{file=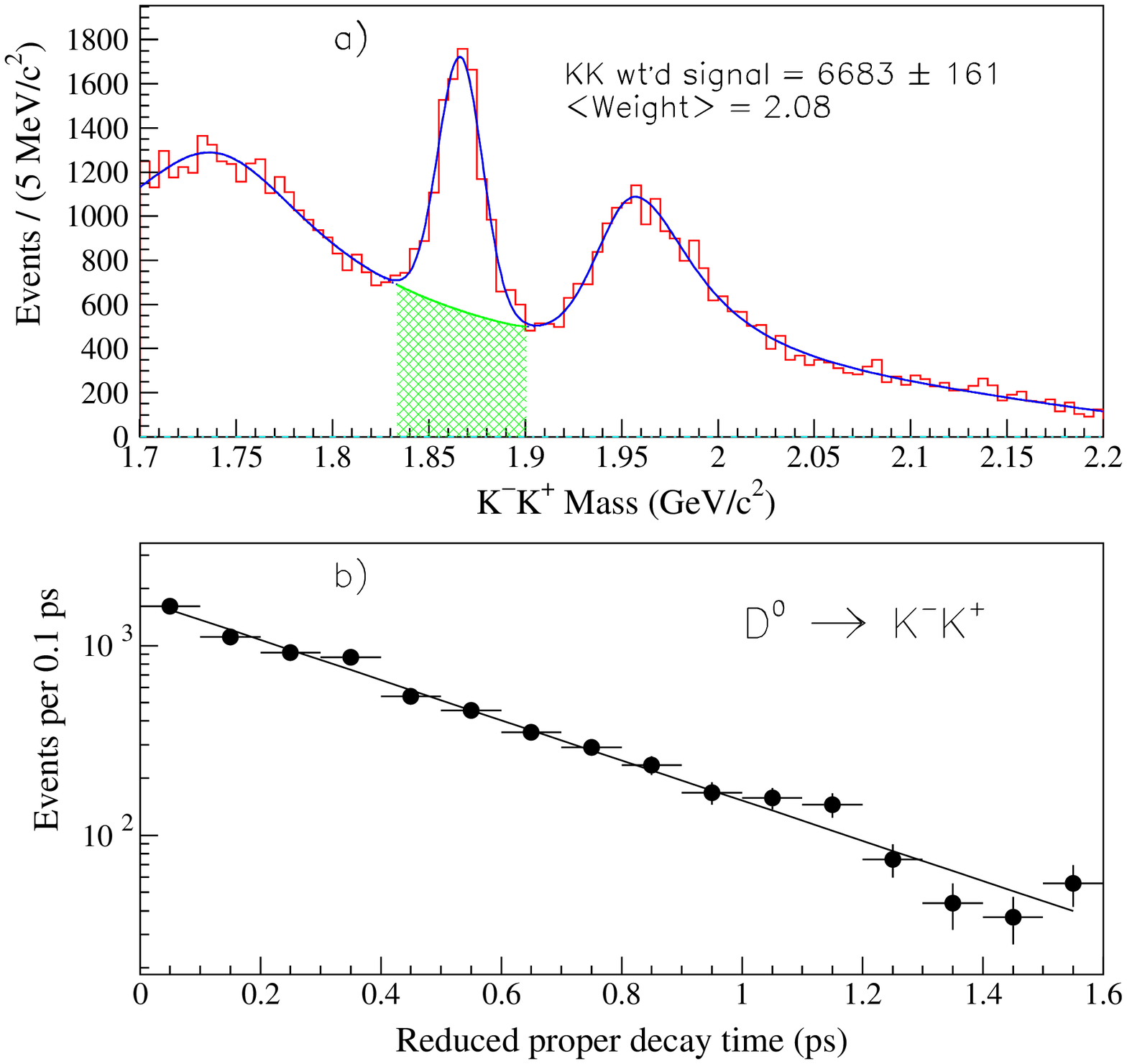,width=6.0cm} 
\caption{E791 \DZ\decays\KM\KP\ (a) mass plot and 
(b) corrected lifetime distribution.  
Their fit to the mass
distribution includes Breit-Wigners to account for reflections
from misidentified \KM\PIP\ and \KM\PIP\PIZ\ decays.  The dashed
line in (b) shows the results of their lifetime fit.
}\label{e791_dgamma}}
They get roughly equal numbers of events
in the electron mode (\DZ\decays\KM\EP\NE).  They
are employing a ``blind'' analysis --- they will not look in the wrong
sign signal region until their cuts have been optimized and their backgrounds
are understood.  They will extract $r_{\rm mix}$ from a two dimensional fit 
to $\Delta m$ and $t$.  They predict a sensitivity by extrapolating 
from preliminary studies:  
if they observe precisely zero background subtracted 
events in their wrong sign signal region, their 90\% CL limit on 
$r_{\rm mix}$ 
will be 0.12\%.  

FOCUS does not yet have a prediction for the  
sensitivity of their hadronic mode search.  They have
a large sample of reconstructed \DS\ tagged \DZ decays, with 
approximately 150,000 \DZ\decays\KM\PIP,
\DZ\decays\KM\PIP\PIM\PIP, and \DZ\decays\KM\PIP\PIZ\ candidates.
With this large sample and excellent proper time resolution 
they should be competitive with CLEO.

\section{Lifetime Difference Searches}

E791 has published 
results of a search for $\Delta \Gamma$, while  
CLEO and FOCUS searches are in progress.

\subsection{E791 $\Delta \Gamma$ Search}

E791 searched~\cite{e791_dgamma}\ for
a lifetime difference between the CP$^+$ and CP$^-$ eigenstates of the
\DZ.  To do so, they compared the lifetimes
of the decays 
\DZ\decays\KM\KP\ (CP$^+$) and \DZ\decays\KM\PIP\ 
($1\over2$ CP$^+$ + $1\over2$ CP$^-$):
\begin{eqnarray}
{{\Gamma(\KM\KP) - \Gamma(\KM\PIP)}\over{\Gamma(\KM\PIP)}} & = & 
{{\Gamma_+ - {\textstyle {1\over2}}(\Gamma_+ + \Gamma_-)}\over{\Gamma(\KM\PIP)}} \nonumber \\ 
& = & 
y_{\rm CP}.
\label{dgammahalf}\end{eqnarray}
\FIGURE{
\epsfig{file=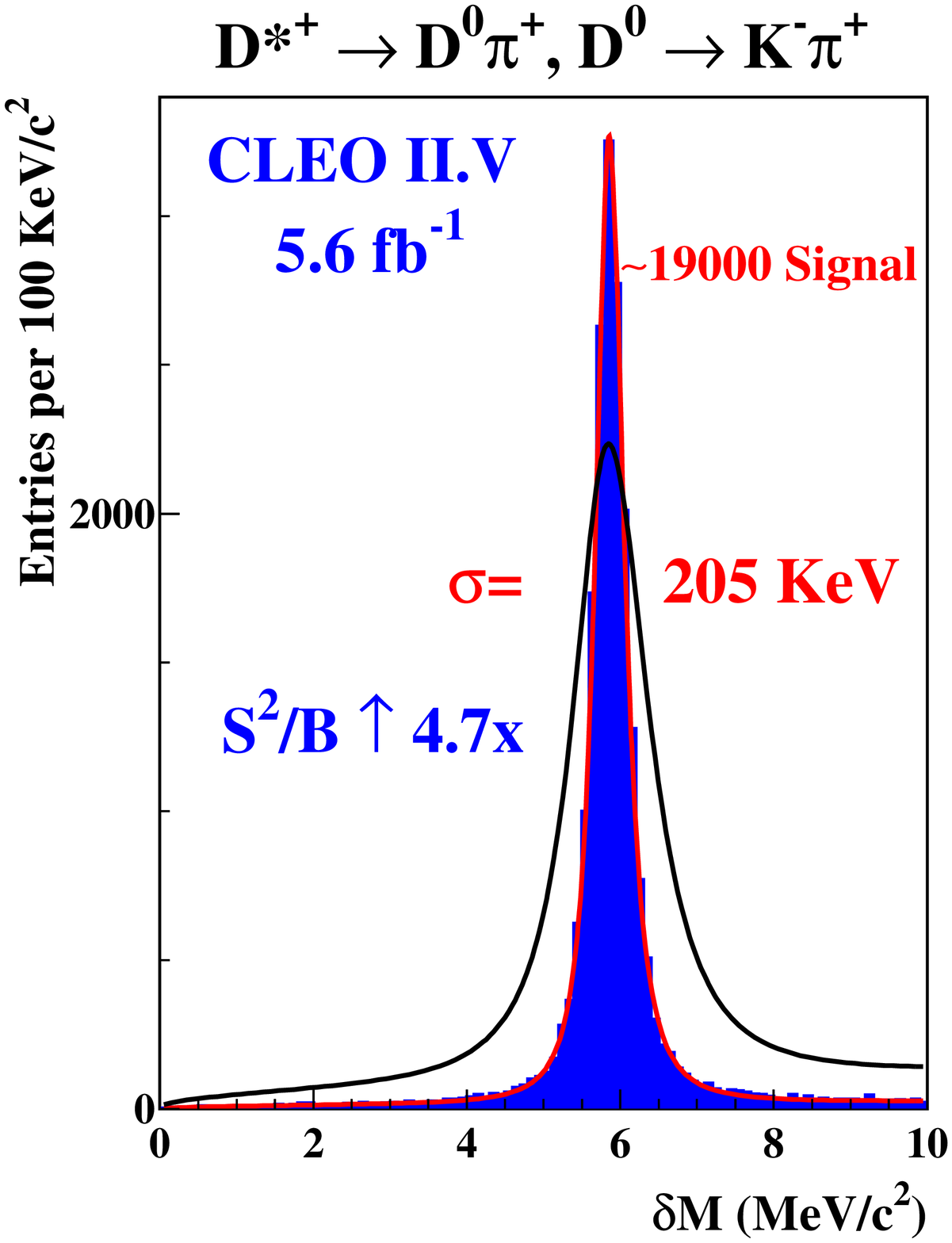,width=2.7cm}\hspace*{0.4cm} 
\epsfig{file=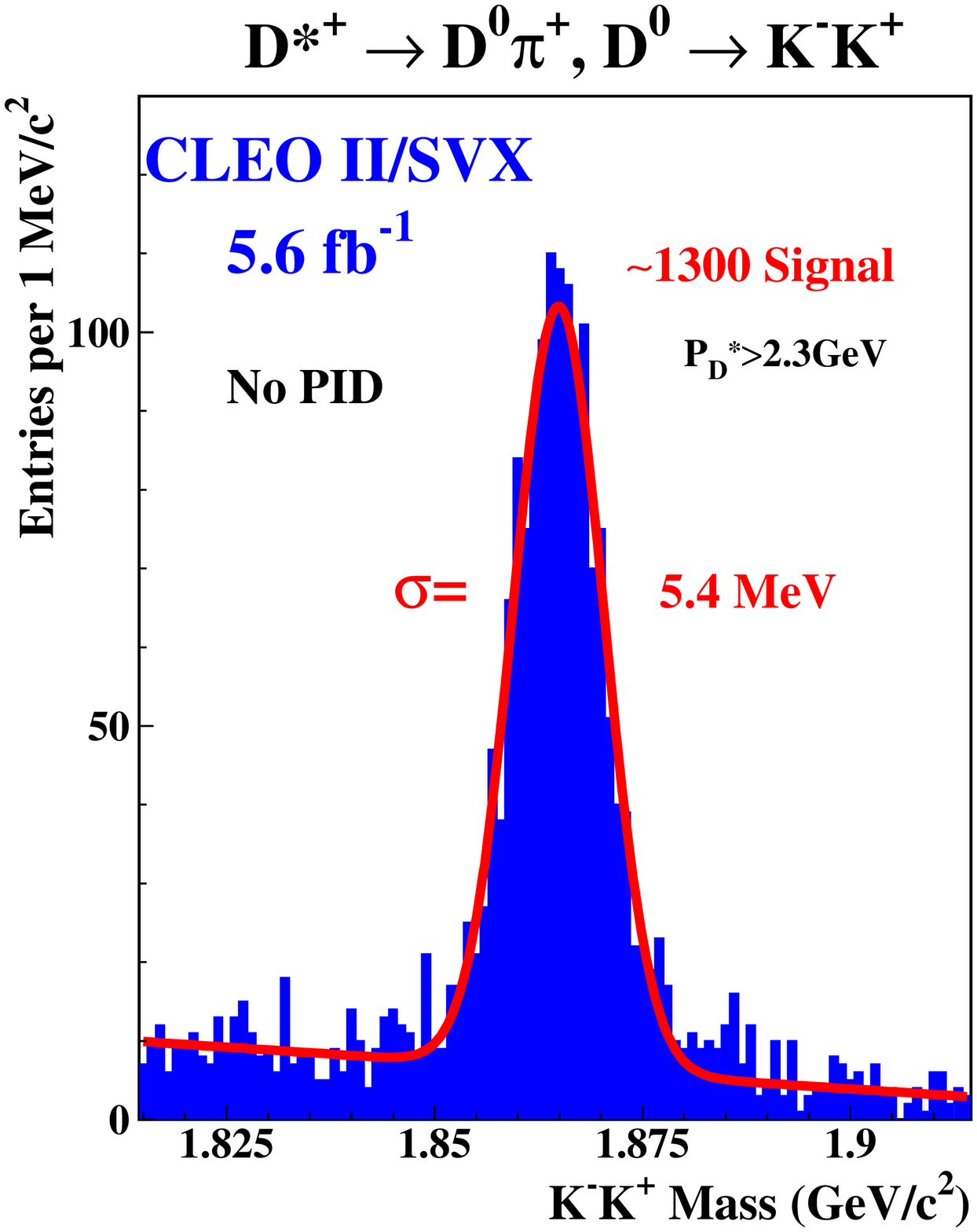,width=2.8cm} 
\caption{CLEO \DZ\decays\KM\PIP\ and 
\DZ\decays\KM\KP\ mass plots, for candidates used in their
preliminary $\Delta \Gamma$ search.
}\label{cleo_dgamma}}
The mass plot for the \KM\KP\ candidates used in their analysis is
shown in figure \ref{e791_dgamma}, along with the reduced proper time
distribution used to extract the lifetime for this decay mode.
Their fit to the mass
distribution includes Breit-Wigners to account for reflections
from misidentified \KM\PIP\ and \KM\PIP\PIZ\ decays.  The dashed
line in plot (b) of the figure shows the results of their lifetime 
fit\footnote{If $\Delta \Gamma \neq 0$,  the proper time 
distribution for $K\pi$ is not strictly exponential.  Given the previous 
limits on $\Delta \Gamma$, however, E791 notes~\cite{e791_dgamma}\ 
 that an exponential
is an excellent approximation given their resolution.}.
After cuts, they have $3213 \pm 77$ signal events in \KM\KP\ 
and $35,427 \pm 206$ in \KM\PIP.  
They observe no difference in lifetimes, and
quote:
\begin{eqnarray}
y_{\rm CP} = 0.008 \pm0.029\pm0.010\hspace*{0.3cm} \nonumber \\
-0.04 < y_{\rm CP} < 0.06\ \ \ \  (90\% {\rm CL}).
\label{e791_dg_results}\end{eqnarray}

\subsection{CLEO $\Delta \Gamma$ Search}

CLEO has shown very 
preliminary results on $\Delta \Gamma$
based on 5.6~fb$^{-1}$ of the 9.1~fb$^{-1}$ data sample they have
collected with the CLEO II.V detector.  They compare the lifetimes of 
\DZ\decays\KM\KP\, \DZ\decays\PIM\PIP (both CP$^+$) and 
\DZ\decays\KM\PIP\ to extract a measurement of $y_{\rm CP}$ via 
equation \ref{dgammahalf}.

They measure the proper time of decays using only $y$ coordinate
information: 
\begin{displaymath}
\tau_{\DZ} = 
{{y_{\rm vtx} - y_{\rm beamspot}}\over{c}}\cdot{{m_{\DZ}}\over{p_y}}
\end{displaymath}
(their beam spot has $\sigma_y \approx 10 \mu m$, 
$\sigma_x \approx 250 \mu m$).

Based on samples of roughly 
$1300$ \KM\KP, $475$ \PIM\PIP, and $19000$ \KM\PIP\ 
(figure \ref{cleo_dgamma}) events they find:
\begin{eqnarray}
y_{\rm CP} = 0.032\pm0.034\pm0.008\hspace*{0.4cm} \nonumber \\
-0.076 < y_{\rm CP} < 0.012\ \ \ \  (90\% {\rm CL}).
\label{cleo_dg_results}\end{eqnarray}
With much more data to analyze, they believe their sensitivity will
increase substantially.

\FIGURE{
\epsfig{file=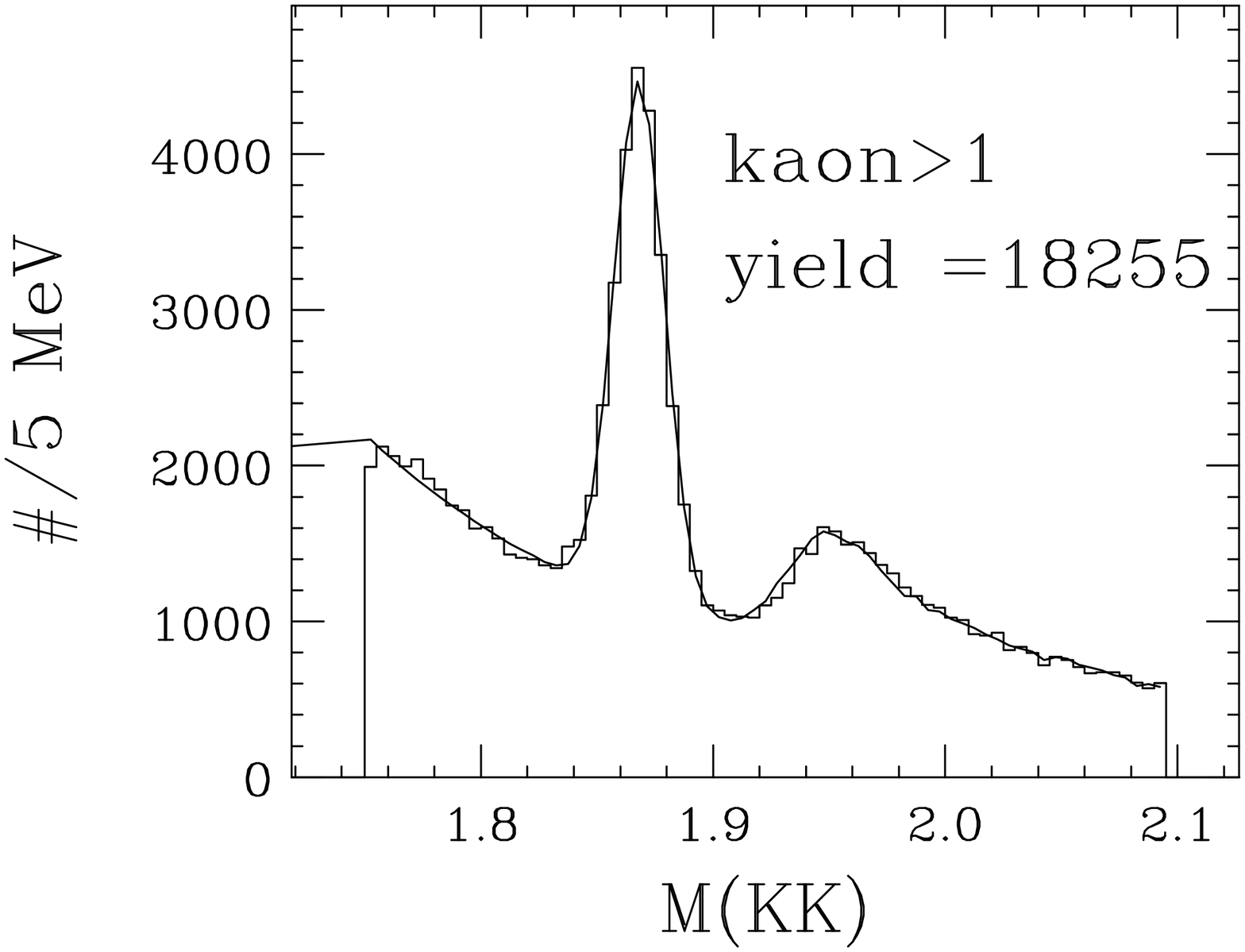,width=3.3cm}\hspace*{0.35cm} 
\epsfig{file=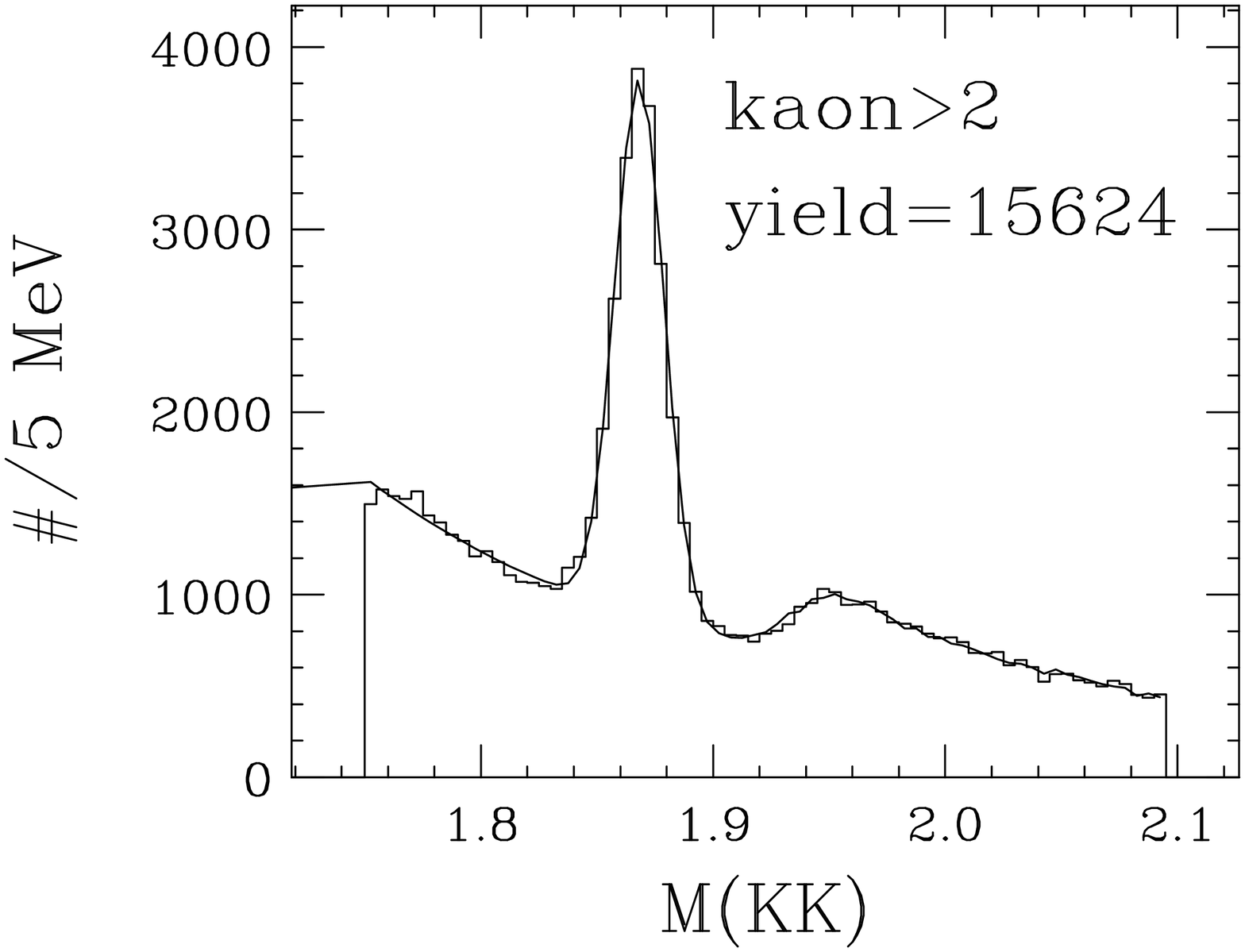,width=3.3cm}\caption{FOCUS 
\DZ\decays\KM\KP\ mass plots. The $\KM\PIP$ reflection
can be controlled and understood by varying the kaon 
identification requirement (which is tighter in the plot on the right).  
}\label{e687_dgamma}}

\subsection{FOCUS $\Delta \Gamma$ Search}

FOCUS will also 
search for $\Delta \Gamma$ by comparing the lifetimes of the
CP$^+$ final states $\KP\KM$ and $\PIP\PIM$ to that of $\KM\PIP$ (equation 
\ref{dgammahalf}).  Mass plots for their $\KP\KM$ candidates are shown in
figure \ref{e687_dgamma}.  By varying the $K$ identification cuts, they
can control and study the reflection from $\KM\PIP$ observable at 
$\approx 1.96$~GeV/c$^2$.  This reflection does not overlap
with the $\KP\KM$ signal, and FOCUS includes a contribution from this
reflection in its fits.  Due to excellent proper time resolution
($\approx7$\% of the \DZ\ lifetime) the fractional error in their
lifetime is equal to the fractional error in their $\KP\KM$ yield.
They expect a statistical error on $y_{\rm CP}$ of 
$0.011-0.013$. 

\section{Rare Decay Searches}

Just as for mixing, the GIM mechanism leads to a strong suppression
of FCNC decays such as 
$\DZ\decays\EPEM$ or $\DP\decays\PIP\MP\MM$, which have expected branching
ratios in the range 
$10^{-19} - 10^{-6}$~\cite{Pakvasa,Schwartz}.  Dilepton modes such as these 
have experimental advantages which make them well suited for sensitive 
searches: lepton
identification is typically clean and efficient, and leptons are
infrequent in charm events so that combinatoric backgrounds
are minimal.  

Dilepton modes that violate lepton flavor number
(LFNV) or lepton number (LNV) such as $\DP\decays\PIP\MP\EM$ and
$\DP\decays\PIM\EP\EP$ would be clear evidence for new physics and
experiments typically look for these decays in addition to FCNC modes.

\FIGURE{
\epsfig{file=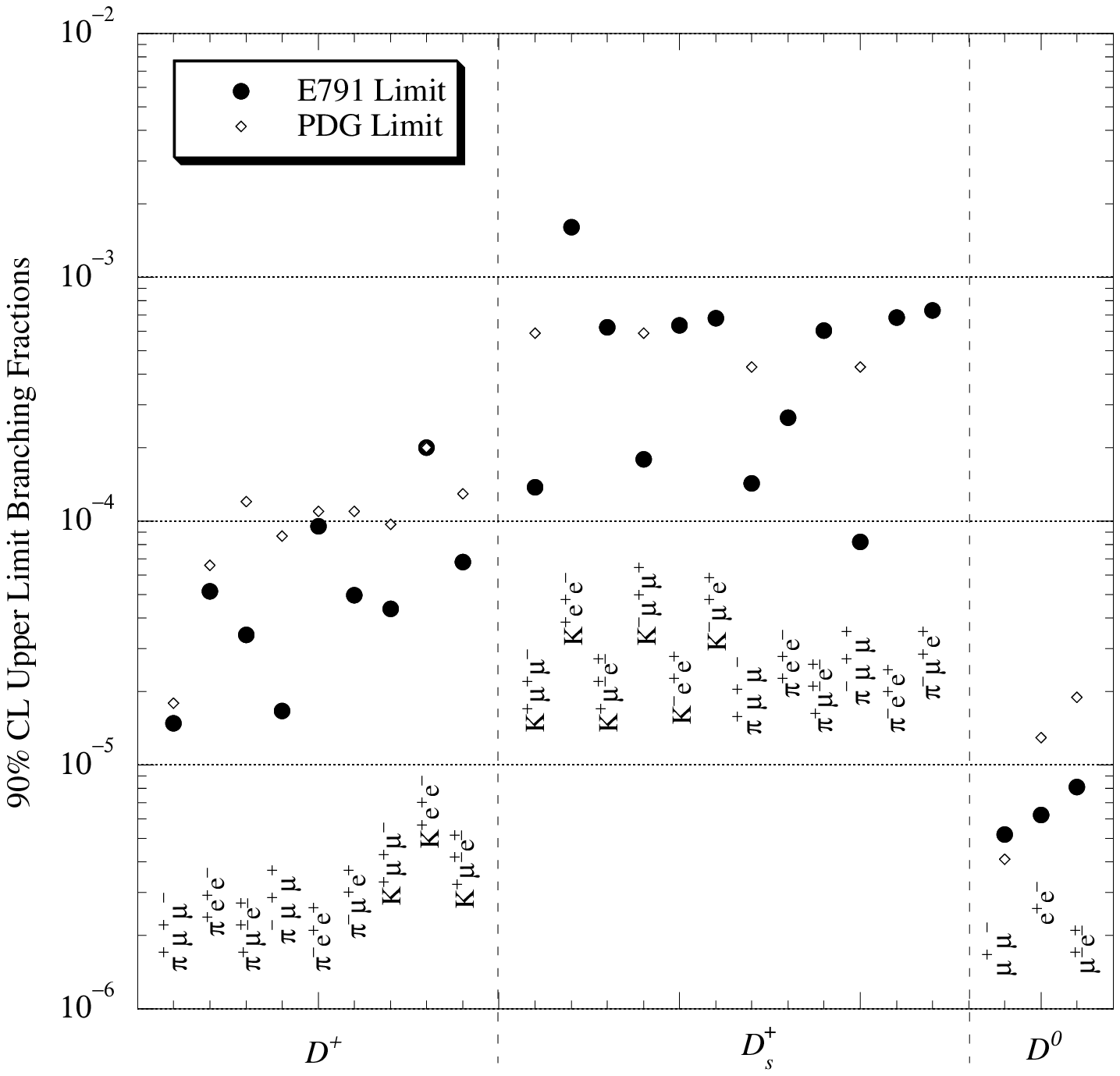,width=6.0cm}\caption{Comparison
of the 90\% CL upper-limit branching fractions from E791 data (dark 
circles) with previous best limits (open diamonds).
}\label{e791_rarefig}}
E791 has recently published results on searches for 
several FCNC, LFNV, and LNV modes.  FOCUS has an analysis in progress
and has shown predicted sensitivities.

\subsection{E791 Rare Decay Search}

E791 performed a blind search~\cite{e791_rare} for 24 rare and forbidden
modes.  
They find no evidence for any of them and set 90\% CL upper
limits on the number of events ($n^*_{X\ell\ell}$) they observe using
the prescription of Feldman and Cousins~\cite{feldman}.  They turn 
this into an upper limit on the branching ratio for each mode by using
a normalizing mode:
\begin{equation}
B(X\ell\ell) = {{n^*_{X\ell\ell}}\over{n_{\rm norm}}} \cdot 
{{\epsilon_{\rm norm}}\over{\epsilon_{X\ell\ell}}} \cdot
B({\rm norm})
\label{rareBR}\end{equation}
where $\epsilon$ is detection efficiency.  
An important aspect of this method is the comparison to a normalizing
mode.  If this mode is similar to the mode of interest,
many sources
of systematic error cancel due to the ratio of detection efficiencies
in equation \ref{rareBR}.
For example, for rare \DP\ decay modes,   
E791 uses $\DP\decays\KM\PIP\PIP$.
The branching fraction $B_{\rm norm}$ is taken from the Particle
Data Book~\cite{PDG}. 
They incorporate systematic errors into their limits using the methods
of Cousins and Highland~\cite{Cousins}.  

The results of their searches are compared to previous limits in 
figure~\ref{e791_rarefig}.  In nearly all cases E791 is the first to
report a limit or has significantly improved the previous limit.

\subsection{FOCUS Rare Decay Search}

The FOCUS experiment is in the early stages of its search for rare
and forbidden dilepton modes.  They are also doing a blind search,
and so can only quote expected ``sensitivities,'' which they
define as the limits they would
set if the number of events in the signal region equals the 
predicted number of 
background events.  Their analysis is very similar to that
performed by E791.  Currently their sensitivities for \DP\ modes
are at the $4-6\times10^{-6}$ level, although this may  improve.

\section{Summary}

It is very difficult to summarize the recent activity in experimental
searches for charm mixing: there is little overlap in the assumptions
made by experiments that have recently shown results.  The best limits
in the 
most general case 
(minimal assumptions) come from E791:
\begin{eqnarray*}
r_{\rm mix}(\DZ\decays\DZB) &<& 1.45\%.
\nonumber \\
\overline{r}_{\rm mix}(\DZB\decays\DZ) &<& 0.74\%
\end{eqnarray*}
If one assumes $|q/p|=1$, E791 also has the best published limit:
\begin{displaymath}
r_{\rm mix} < 0.5\%. 
\end{displaymath}
from a search using semileptonic decays.

%
Analyses in progress by CLEO and FOCUS are sensitive to 
$r_{\rm mix} \sim 5\times10^{-4}$ (maybe lower).

First results from searches for a 
lifetime difference $\Delta \Gamma$ have recently
been published by E791.  They find 
$-0.04 < y_{\rm CP} < 0.06$ at 90\% CL.  Since $D_H$ and $D_L$
are at least approximate CP eigenstates,
$y_{\rm CP} \approx y$.  
CLEO and FOCUS will push limits on $y_{\rm CP}$
to the few $\times$ $10^{-3}$ level in the very near future.

E791 has recently published limits on the branching ratios for 
24 rare and forbidden decay modes.
In nearly all cases their results   
are new or are a significant improvement over previous
limits.  Their limits on the branching ratios for 
\DP\ rare decays, for example, are at the level 
of a few $\times$ $10^{-5}$.  The rare decay search of the FOCUS 
collaboration is in progress, and they are 
sensitive to \DP\ branching ratios of a few 
$\times$ $10^{-6}$.

It is interesting to note that $\DP\decays\PIP\MM\MP$ can occur via 
$\DP\decays\PIP\phi\decays\PIP(\MM\MP)$  The branching ratio for this decay
is $1.5\times10^{-6}$, which can be calculated from the known branching ratios
for $\DP\decays\PIP\phi$ ($6.1\times10^{-3}$) and
for $\phi\decays\MP\MM$ ($2.5\times10^{-4}$)~\cite{PDG}.  
Experiments are getting 
tantalizingly close to seeing this, which could be used as a ``calibration''
point for experimental sensitivity.



\acknowledgments
I would like to thank the local conference committee for organizing 
an interesting and productive meeting.
Thanks also go to those members of the ALEPH, CLEO, E791, and FOCUS 
collaborations who provided me with information and plots relevant to 
their searches for charm mixing and rare decays.

\end{document}